# Quasi-Quantum Model of Potentization

MARCIN MOLSKI[*]

Department of Theoretical Chemistry, Faculty of Chemistry
Adam Mickiewicz University of Poznań
ul. Grunwaldzka 6, PL 60-780 Poznań, Poland

## Abstract

Analytical time-dependent functions describing the change of the concentration of the solvent S(t) and the homeopathic active substance A(t) during the decimal and centesimal dilution are derived. The function S(t) is a special case of the West-Brown-Enquist curve describing the ontogenic growth, hence the increase in concentration of the solvent during potentization resembles the growth of biological systems. It is proven that the macroscopic S(t) function is the ground state solution of the microscopic non-local Horodecki-Feinberg equation for the time-dependent Hulthèn potential at the critical screening. In consequence the potentization belongs to the class of quasi-quantum phenomena playing an important role both in the biological systems and homeopathy. A comparison of the model proposed with recently performed experiment on delayed luminescence of the homeopathic remedy will be also made.

[*] Corresponding author; fax: +48618658008, e-mail: MAMOLSKI@AMU.EDU.PL

# Introduction

Homeopathic remedies are micro-dosed substances derived from botanical, animal or mineral sources by a successive dilution and vigorously shaking (succussion) referred to as *potentization*. This process converts the original substance into a therapeutically active medicine of decimal (1/10), centesimal (1/100) or quinquagenimillesimal (1/50 000) dilution rate. One usually uses steps of 1/10 for the spreading out of substance in a medium, both when dealing with liquids and solid, insoluble substances that are potentized in a mortar. The homogenization of the active substance involves (when liquids are concerned) shaking whereas the homogenization of solid substances is obtained through grinding. The optimal rhythm for the movement of the liquid is of oscillatory type, which causes the optimal homogenity of the substance in the solution. The dilution attained after one decimal dilution is termed $D_1$ and is used as the starting point for preparing the next dilution $D_2$ in exactly the same way as before. This process including spreading out of substance, rhythmic dispersion (homogenization) can be continued indefinitely even beyond Avogadro number, when no molecules of the active substance are present in remedy. To explain the homeopathic activity at high dilutions the researchers have been used the theoretical models which refer to water polymers[1], clathrates[2],electric dipoles[3], vortices[4] and other[5-8] mechanisms and structures assumed to be carriers of information transferred from the molecules of the active substance to the ordered molecules of the solvent produced by potentization. According to the specified models[1-8] this information is administered during a homeopathic treatment.

A careful reading of the Hahnemann's *Organon of Medicine*[9] reveals that he believed in the possibility of exciting in the homeopathic remedy a *spirit-like* power of medicines (*fast geistige Kraft der Arzneien*) or *a vital principle of animal life* or *vital force* by making use of the potentization. According to Hahnemann this *vital force* is immanent component of the homeopathic remedy besides the solvent (water, ethanol, lactose) and active substance employed in its production. It is interesting that the concept of *spirit-like* power of homeopathic drugs has been completely abandoned in contemporary homeopathy as it has not been supported by theoretical models and experimental evidence. The main objective of the present work is proving that the concentration of the solvent in which the active homeopathic substance is diluted increases according to the function being a special case of the West-Brown-Enquist curve describing the ontogenic growth. Hence, the remedy prepared according to the orthodox homeopathic rules is endowed with a *vital force* or *power to growth* - the same as a growing biological system. A comparison of the model proposed with recently performed experiment on delayed luminescence of the homeopathic remedy will be also made.

# Mathematics of potentization

Let's assume that the active substance of mass $m_A$ is dissolved and homogenized in the solvent of mass $m_S$ by making use of the succussion procedure and decimal dilution. In such circumstances at every step of potentization the following relationship is satisfied

$$m_A + m_S = M$$

in which M=const is the total mass of the remedy prepared. For example, $D_1$ and $D_2$ potencies can be described by the formulae (the unit mass is gram)

$$D_1: \quad \frac{1\,[g]+9\,[g]}{10} = 0.1\,[g]+0.9\,[g]=1\,[g]$$

$$D_2: \quad \frac{\frac{1\,[g]+9\,[g]}{10}+9\,[g]}{10} = 0.01\,[g]+0.99\,[g]=1\,[g]$$

In the similar manner one may produce remaining potencies using both decimal and centesimal dilution. The results are presented in Table 1.

Dividing the mass relation by M one gets

$$A(x)+S(x)=1$$

in which $A(x)=m_A/M$ and $S(x)=m_S/M$ multiply by 100% denote the concentration of the molecules of the active substance A(x) and solvent S(x) in remedy, whereas x=1,2,3..N stands for the step of the potentization. The functions A(x) and S(x) can be interpreted in an alternative manner as the probability of finding a molecule of the active substance or a molecule of the solvent in the homeopathic remedy. Analysing the results presented in Table 1 one may prove that for the decimal dilution A(x) and S(x) can be specified explicitly as

$$A(x)=10^{-x} \qquad S(x)=1-10^{-x}$$

whereas for the centesimal solution we have

$$A(x)=100^{-x}=10^{-2x} \qquad S(x)=1-100^{-x}=1-10^{-2x}$$

# Potentization time

According to Hahnemann[9] the potentization (dynamization) of the medicine is obtained by a precise numbers of shakings and dilutions in the given time sequences, or by an exact number of mixing (triturations) of a diluted medicinal substance. For example, the dispersion and homogenization of the active substance in the liquid solvent take usually place for 4 minutes for mineral substances and 2.5 minutes for plant substances and animal compounds whereas the homogenization of solid substances through grinding takes about 1 hour in every step. Those facts indicate that the each potentization is performed in the same time interval. Hence, we can introduce a *potentization time* $t_0$, which is indispensable to produce $D_{x+1}$ dilution from the $D_x$ diluted active substance. In other words the potentization step can be given in the form of the time-dependent function

$$\frac{t}{t_0} = x(t)$$

Now, one may express the x-dependent functions A(x) and S(x) in the time-dependent form

$$A(t) = 10^{-x(t)} = \exp(-at) \qquad S(x) = 10^{-x(t)} = 1 - \exp(-at)$$

$$a = \frac{\ln(10)}{t_0} = \frac{2.302585093}{t_0}$$

which describe the decrease of the active substance in the solvent and increase of the solvent concentration in the remedy. Applying the same mathematical procedure for centesimal dilutions one gets the formulae

$$a = \frac{\ln(100)}{t_0} = \frac{4.605170186}{t_0}$$

The functions A(t) and S(t) precisely determine the concentration of the active substance and the solvent in the remedy at the each step of the potentization procedure performed at the time $t_0$ including succussion and dilution.

## The first- and second-order dynamization

The function describing the increase in the solvent concentration during potentization satisfies the first- and second-order differential equations

$$\frac{d}{dt} S(t) - a \frac{\exp(-at)}{1 - \exp(-at)} S(t) = 0$$

$$\frac{d^2}{dt^2} S(t) + a^2 \frac{\exp(-at)}{1 - \exp(-at)} S(t) = 0$$

The second term in the above equation represents the well known in the quantum physics Hulthèn potential[2] widely used in description of the electrostatic interactions between micro-particles. The above equation can be expressed in the dimensionless coordinate τ=at

$$\frac{d^2}{d\tau^2} S(\tau) + \frac{\exp(-\tau)}{1 - \exp(-\tau)} S(\tau) = 0$$

One may prove that the above equation is a special case of the quantal non-local Horodecki-Feinberg equation[10,11] for the time-dependent Hulthèn potential[12] at the *critical screening*[13] (see Appendix).

This result indicates that the process of increasing concentration of the molecules of the solvent during preparation of the homeopathic medicine belongs to the class of quasi-quantum phenomena. The notion *quasi-quantum* refers to the possibility of application of the quantum language and formalism in description of macroscopic phenomena like potentization process. In particular the second-order kinetic equation governing the potentization takes identical form as the microscopic eigenvalue equation for the quantized eigenstate equal to zero. Since this equation is a special case of the non-local Horodecki-Feinberg equation, the potentization belongs to the class of macroscopic non-local phenomena. The macroscopic second-order equation does not contain the Planck's constant whereas its eigenfunction S(τ) can be interpreted in probabilistic terms as the probability of finding a molecule of the active substance in the homeopathic remedy. Hence, it is consistent with the predictions of the weak quantum theory developed by Atmanspacher and coworkers[14].

## Ontogenic growth and potentization

The function S(τ), which describes the concentration or the solvent in the remedy at the each step of the potentization is well known in the biological domain. In 2001 West, Brown and Enquist[15] (WBE) formulated a general model for ontogenic growth from the first principles. On the basis of the conservation of metabolic energy, the allometric scaling of metabolic rate, and energetic costs of producing and maintaining biomas, they derived the function

$$m(t) = M\left[1 - c\exp(-c_1 t)\right]^{\frac{1}{c_2}}$$

$$c = 1 - \left(\frac{m_0}{M}\right)^{\frac{1}{4}} \quad c_1 = -\frac{a}{4M^{1/4}} \quad c_2 = \frac{1}{4} \quad m(t=0) = m_0 \quad m(t=\infty) = M$$

which fits very well the data for a variety of different species from protozoa to mammalians. Here, $m_0$ is the initial mass of the system whereas M denotes the maximum body size reached. The WBE function can be expressed in dimensionless time-coordinate

$$r(\tau) = \left[1 - \exp(-\tau)\right] \qquad r(\tau) = \left(\frac{m_0}{M}\right)^{c_2}$$

in which

$$\tau \;=\; c_1(t - t_e) \qquad t_e = \ln(c)/c_1$$

As it has been proved by WBE[15], the above function provides the powerful way of plotting the data that reveals universal properties of biological growth. If the mass ratio is plotted r(τ) against a variable τ then all species (mammals, birds, fish, crustacea), regardless of taxon, cellular metabolic rate and mature body size M fall on the same parameterless universal curve r(τ).

A comparison of the universal WBE function with the homeopathic function S(τ) reveals that the latter can be obtained from the former by the substitution

$$m_0 \to 0 \qquad c_2 \to 1$$

The condition $m_0=0$ reflects the fact that the initial mass of the solvent during potentization is equal to zero, whereas $c_2=1$ indicates the different mass scaling of this process in comparison with the mass scaling of the biological growth ($c_2=3/4$).

## Conclusions

From the scientific point of view the potentization seems to be an irrational and misterious procedure, which is difficult to explain by the well established physical theories. The results obtained in this work indicate that the potentization can be explained in rational terms using the concepts of potentization time and molecular dispersion. When the active substance is diluted in the solvent and then vigorously shaken by strikes in a succussion procedure then two processes on the micro-level take place: first – molecular dispersion of the conglomerates of the substance, and second – removal of the active molecules off the tincture in the series of dilutions. From the physical point of view, the precise numbers of shakings and dilutions in the given time sequences, or by an exact number of mixing (triturations) of a diluted medicinal substance is a kind of the *homeopathic clock*, which permits introduction of the *potentization time* and description of the time-change of the concentration of the solvent. This function is a special case of the WBE function describing the ontogenic growth. If biological growth is characterized by a *power to growth* or a *vital force* then the potentization procedure should excite the same *vital force* in the solvent during preparation of the homeopathic remedy. In other words, the increase in concentration of the molecules of the solvent during potentization resembles growth of biological systems. This conclusion should be treated as a hypothesis, which is consistent with Hahnemann[9]: *.....remarkable transformation of the properties of natural bodies through the mechanical action of trituration and succussion on their particles (while these particles are diffused in an inert dry or liquid substance) develops the latent dynamic powers previously imperceptible and as it were lying hidden asleep in them. These powers electively affect the vital principle of animal life. This process is called dynamization or potentization (development of medicinal power), and it creates what we call dynamizations or potencies of different degrees.*

The second aspect of the potentization – molecular dispersion - is a condition *sine qua none* for the total homogenization of the active substance in the solvent as only then the functions A(t) and S(t) are perfectly satisfied whereas S(τ) has identical form as WBE function describing the ontogenic growth. In such circumstances the remedy prepared according to the orthodox homeopathic rules is endowed with a *vital force* - the same as a growing biological system. In this picture the process of preparation of the homeopathic medicine reproduces the biological growth and excites in the remedy a *spirit-like* power of medicines.

The removal of the active molecules off the tincture in the series of dilutions results in diminishing the mean distance between molecules of the solvent and increasing interactions between them. This effect is connected with increasing value of the Hulthèn potential energy (see Fig.1) of the solvent justifying the homeopathic terms: *potentization* and *dynamization* of the homeopathic remedy during succession and dilution.

Because the macroscopic function S(τ) describing the change in concentration of the solvent in the remedy is solution of the microscopic non-local Horodecki-Feinberg equation, the potentization procedure belongs to the class of non-local quasi-quantum phenomena. It means that the succussion generates molecules of the solvent in the correlated (quasi-entangled) state amenable to form complex structures against decoherence due to collisions with other molecules, exchanging the electromagnetic radiation and chaotic thermal influences. The results reported by Del Guidice *et al*[3] confirmed that the water molecules can move in highly correlated and ordered way due to interactions between water electric dipole and radiation field, which produce quasi-ordered structures in macroscopic domain. According to Weingärtner[8], such correlated molecular configurations can be effective carriers of information between molecules of the active substance and molecules of the solvent, during preparation of homeopathic medicines. Hence, they can play the role of material carriers of information that is administered during a homeopathic treatment. This interpretation admits homeopathic activity even if no molecules of the active substance are present in remedy. It is consistent with the Collins[16] model assuming that when the active substance dissolved in water becomes more dilute, the remaining molecules clump together to form aggregates of increasing size. Such aggregates endowed with a *vital force* could affect biological systems, hence providing some possible explanation for the effect of a homeopathic activity.

The theoretical results obtained in this work are consistent with the Lenger-Bajpai-Drexel[17] delayed luminescence experiment performed on *Argentum Metallicum* remedy. Delayed luminescence is the phenomenon of photon emission by a complex living system after exposure to whit light for a few seconds. The photon signal is observed after a few milliseconds delay and is observable for a few minutes. The shape of the signal can be theoretically analyzed in terms of four parameters: $t_0$, $B_0$, $B_1$, $B_2$ describing the change in time of the numbers of photons emited

$$n(t) = B_0 + B_1/(t+t_0) + B_2/(t+t_0)^2$$

The coefficients $B_0$ and $B_1$ take significant values in living systems while coefficient $B_2$ gives contribution in non-living complex systems. The delayed luminescence signals of *Argentum Metallicum* were characterized by the coefficient $B_2$ typical for the delayed luminescence of non-living systems, but also by the coefficient $B_0$ typical for living systems. Both coefficients indicate the presence of holistic quantum structures in homeopathic remedy[17] and attribute to it characteristic similar to that observed in the living systems.

# Appendix

The non-local quantum states of a particle of mass *m* moving with superluminal velocity in the field of the time-dependent vector potential *V(t)* is described by the Horodecki-Feinberg equation[10,11]

$$-\frac{\hbar}{2mc^3}\frac{}{dt^2}+\frac{1}{c}V(t)\Psi = P\Psi \tag{1}$$

Here Ψ represents a non-local matter wave associated with the superluminal particle of momentum P, ħ= 1,05457266·$10^{-34}$ J·s is the Planck constant divided by 2π, c is the light velocity. Equation (12) represents the non-relativistic version of the relativistic Feinberg equation for non-local faster than light objects. It has been derived by Horodecki[11] by taking advantage the same procedure as that used in deriving the Schrödinger equation (11) from the relativistic Klein-Gordon equation for local slower than light particles.

The Feinberg-Horodecki equation with the time-dependent Hulthèn potential[12]

$$-\frac{\hbar^2}{2mc^2}\frac{d^2}{dt^2}\Psi_v - V_0\frac{\exp[-c_1(t-t_e)]}{1-\exp[-c_1(t-t_e)]}\Psi_v = Pc\Psi_v \tag{2}$$

can be specified in a dimensionless form[18]

$$\frac{d^2}{d\tau^2}\Psi_v + \beta^2\frac{\exp[-\tau]}{1-\exp[-\tau]}\Psi_v = \varepsilon_v^2\Psi_v \quad \varepsilon_v = \frac{\beta^2 - v^2}{2\beta v} \quad v = 1,2,3... \tag{3}$$

in which

$$\beta^2 = \frac{2mc^2V_0}{\hbar^2} \qquad \varepsilon^2 = -\frac{2mc^3P_v}{\hbar^2} \tag{4}$$

The eigenfunctions of the Horodecki-Feinberg equation take the form[18]

$$\Psi_v = \exp(-\varepsilon\tau)\left[1-\exp(-\tau)\right]_2F_1[2\varepsilon_v+1+v, 1-v, 2\varepsilon_v+1; \exp(-\tau)] \tag{5}$$

For β=1 and ground state v=1 we have ε=0 and $P_1$ =0, whereas the ground state solution $\Psi_1$ reduces to the function S(τ) describing the concentration of the solvent in the homeopathic remedy

$$\Psi_1 = \left[1-\exp(-\tau)\right]\exp(-\varepsilon_1\tau) \xrightarrow{\beta=1, v=1} S(\tau) = 1-\exp(-\tau) \tag{6}$$

This result indicates that the universal growth function can be identified with the ground state solution of the Feinberg-Horodecki equation for the time-dependent Hulthèn oscillator at the *critical screening*[13]. Then β=1 and the momentum eigenvalue is equal to zero $P_1$ =0.

# Acknowledgment

The main idea of this work was presented on the 64th LMHI Congress Warsaw 2009.

Table 1. Decimal and centesimal dilutions of the active substance of concentration A(x) in the solvent of concentration S(x) in which x=1,2….N denotes the potentization step.

| $D_x$ | A(x) | S(x) | $C_x$ | A(x) | S(x) |
|---|---|---|---|---|---|
| $D_1$ | 0.1 | 0.9 | $C_1$ | 0.01 | 0.99 |
| $D_2$ | 0.01 | 0.99 | $C_2$ | 0.0001 | 0.9999 |
| $D_3$ | 0.001 | 0.999 | $C_3$ | 0.000001 | 0.999999 |
| | | | | | |
| $D_n$ | $10^{-n}$ | $1\text{-}10^{-n}$ | $C_n$ | $10^{-2n}$ | $1\text{-}10^{-2n}$ |

Fig.1 The plot of dimensionless S(τ) function and the Hulthèn potential V(τ).

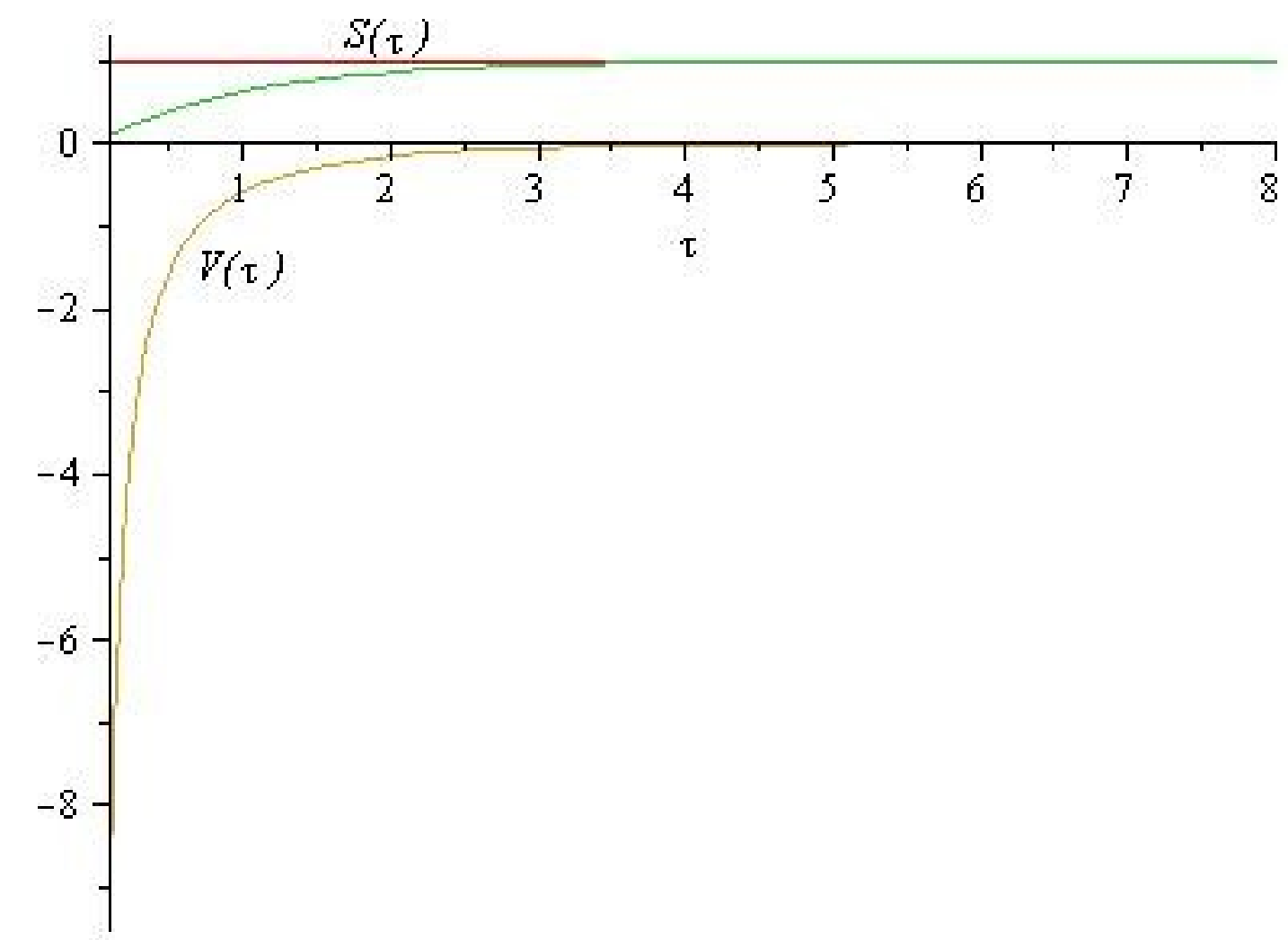